\documentclass[aps,prb,twocolumn,superscriptaddress,citeautoscript]{revtex4}  % for review and submission
\usepackage{graphicx}   % needed for figures
\usepackage{dcolumn}   % needed for some tables
\usepackage{bm}           % for math
\usepackage{amssymb}  % for math
\usepackage{graphicx}
\usepackage{soul, color}
\usepackage{enumerate}
\usepackage{bm}
\usepackage{amssymb}
\usepackage{amsmath}
\usepackage{subfigure, psfrag}
\usepackage{tikz}
\usetikzlibrary{shapes,arrows}
\usepackage{amsmath}
\usepackage{appendix}
\usepackage{multirow, array}
\usepackage{lipsum}

\newcommand*{\citen}[1]{%
  \begingroup
    \romannumeral-`\x % remove space at the beginning of \setcitestyle
    \setcitestyle{numbers}%
    \cite{#1}%
  \endgroup   
}

\begin{document}

\title{Finite-size anomalies of the Drude weight: role of symmetries and ensembles}
%\title{Anomalous transport in spin 1/2 chains I: role of symmetries and ensembles for Drude weight}
\author{R.\ J.\ S\'anchez}
\affiliation{Bethe Center for Theoretical Physics, Universit\"{a}t Bonn, Germany}

\author{V.\ K.\ Varma}
\affiliation{The Abdus Salam ICTP, Strada Costiera 11, 34151, Trieste, Italy}
\affiliation{Initiative for the Theoretical Sciences, The Graduate Center, CUNY, New York, NY 10016, USA}
\affiliation{Department of Engineering Science and Physics, College of Staten Island, CUNY, Staten Island, NY 10314, USA}
\affiliation{Department of Physics and Astronomy, University of Pittsburgh, Pittsburgh, PA 15260, USA}

\date{\today}

\vspace*{-1cm}

\begin{abstract}

We revisit the numerical problem of computing the high temperature spin stiffness, or Drude weight, $D$ of the spin-$1/2$ XXZ chain using exact diagonalization to systematically analyze its dependence on system symmetries and ensemble.  
Within the canonical ensemble and for states with zero total magnetization, we find $D$ vanishes exactly due to spin-inversion symmetry for all but the anisotropies $\tilde \Delta_{MN} = \cos(\pi M /N)$ with $N, M \in \mathbb{Z}^+$ coprimes and $N > M$, provided system sizes $L \ge 2N$, for which states with different spin-inversion signature become degenerate due to the underlying $sl_2$ loop algebra symmetry.
All these loop-algebra degenerate states carry finite currents which we conjecture [based on data from the system sizes and anisotropies $\tilde \Delta_{MN}$ (with $N<L/2$) available to us] 
to dominate the grand-canonical ensemble evaluation of $D$ in the thermodynamic limit.
Including a magnetic flux not only breaks spin-inversion in the zero magnetization sector but also lifts the loop-algebra degeneracies in all symmetry sectors --- this effect is more pertinent at smaller $\Delta$ due to the larger contributions to $D$ coming from the low-magnetization sectors which are more sensitive to the system's symmetries.
Thus we generically find a finite $D$ for fluxed rings and arbitrary  $0<\Delta<1$ in both ensembles. 
In contrast, at the isotropic point and in the gapped phase ($\Delta \ge 1$) $D$ is found to vanish in the thermodynamic limit, independent of symmetry or ensemble.
Our analysis demonstrates how convergence to the thermodynamic limit within the gapless phase ($\Delta < 1$) may be accelerated and the finite-size anomalies overcome: 
$D$ extrapolates nicely in the thermodynamic limit to either the recently computed lower-bound or the Thermodynamic Bethe Ansatz result provided both spin-inversion is broken and the additional degeneracies at the $\tilde \Delta_{MN}$ anisotropies are lifted.
\end{abstract}

%\pacs{05.30.Rt, 05.30.Jp, 67.25.D-, 03.75.Lm}
\maketitle
\section{Introduction}
Ideal conduction is observed when the mean free path of particles exceeds the sample size and no scattering events occur.
A key distinguishing feature between such ideal metals and other metals or insulators is their response to external dc fields: unimpeded acceleration of particles in the former produces a delta function in 
its conductivity with a certain spectral weight $D$, the Drude weight.
A particularly convenient way to compute this quantity is through the response of the system's ground~\cite{Kohn64} or excited state~\citep{CastellaZotos,ZotosPrelovsek,PhysRevB.55.11029} eigenfunctions to 
twisted boundary conditions: only ideal metals will produce a finite response, $D>0$, to such a perturbation.
While such \textit{ballistic transport} is generically expected in integrable systems~\citep{Mazur_ineq}, computing $D$ even for simple models can often be done 
only numerically~\citep{MillisAndrei, CabraHonecker, PhysRevB.77.245131, HerbrychZotos, PhysRevLett.108.227206, PhysRevB.87.245128, PhysRevLett.112.120601, VarmaSanchez, Karrasch2017, PhysRevB.94.201112}. 

In this work we compute the spin stiffness or Drude weight $D$ for the paradigmatic integrable quantum Heisenberg chain at high temperatures, for various combinations of the conserved symmetries and ensembles chosen. The important role of the model's discrete and dynamical symmetries in the numeric computation of $D$ has been emphasized before~\citep{MillisAndrei, PhysRevB.77.245131}.
We were motivated to revisit this long-studied problem due to certain inexplicable ensemble-dependent convergence rates of the finite system data to the thermodynamic limit found, for certain interaction strengths, 
in some of our own personal calculations as well as those in previous literature~\citep{CabraHonecker, HerbrychZotos, PhysRevB.87.245128}; the details will be presented along the way.
We attempt to resolve these issues by systematically enlisting the contributions of the current carrying states in each symmetry sector. 
In so doing we find what ensemble and what system symmetries (parity, spin-inversion, $sl_2$-loop symmetry) best help to achieve quickest convergence to the thermodynamic limit, and explicate why. 
We found this last point particularly relevant and revelatory in light of the recent lower bounds for the high-temperature spin Drude weight~\citep{ProsenI} and the particle-based hydrodynamic result~\citep{Ilievski} reported for the XXZ and related models.
 
At inverse temperature $\beta$, and in the absence of any singular long-time behavior of the current-current correlation function $\langle \hat j(t) \hat j(0) \rangle$ and Meisner weight, 
the Drude weight $D$ has the spectral representation~\citep{PhysRevB.55.11029, PhysRevB.77.245131} 

\begin{equation}
D(\beta)  = \pi \beta \sum\limits_n \frac{e^{-\beta E_n}}{L Z}\sum\limits_{m, E_m = E_n} |\langle n |  \hat j | m \rangle |^2,
\label{eq:Drude2}
\end{equation}

\noindent 
where $E_i$ is the eigenvalue associated to the eigenstate $| i \rangle$ of the $L$-site model and $Z$ is the corresponding partition function.
Let us remark here that we have explicitly confirmed the absence of any long-time plateau in the time-dependent component of the autocorrelation function $\langle \hat j(t) \hat j(0) \rangle$ for the finite systems we consider, which may arise from almost-degenerate states ignored in the equation above. Equation~\eqref{eq:Drude2} thus yields the long-time asymptotic value of the current-current correlation function.

In the following we shall numerically evaluate Eq.~\eqref{eq:Drude2} for the spin-$1/2$ XXZ chain in the high-temperature limit, where the meaningful quantity is 
$\lim_{\beta \rightarrow 0} \beta^{-1} D(\beta)$. 
Finite frequency transport anomalies in this system at high temperature is reported elsewhere~\cite{to.be.published}.

\section{Model and symmetries} 

The spin-$1/2$ XXZ model with anisotropy $\Delta \ge 0$ and exchange integral $J$, for a $L$-site spin chain with periodic boundary conditions is described by the Hamiltonian 

\begin{equation}
\hat H = \sum\limits_{j =1}^L \frac{J}{2} \left( \hat S_j^{+} \hat S_{j+1}^{-} + \hat S_j^{-} \hat S_{j+1}^{+} \right) + \Delta \, \hat S_j^z \hat S_{j+1}^z,
\label{eq:Hamiltonian}
\end{equation}

\noindent
where the ladder spin operators ($\hat S_j^{+} \hat S_{j+1}^{-} + \hat S_j^{-} \hat S_{j+1}^{+}$) flip pairs of spins at sites $j, j+1$. 

The model is invariant under spin rotations $ \hat m_z$ about the $z$-axis, lattice translations $\hat T$ and two discrete symmetries, namely (i) spin-inversion $\hat Z$ with eigenvalues $z=\pm1$, defined such that 

\begin{equation}
\hat Z |S_1^z, S_2^z, \, ..., \, S_L^z \rangle = | -S_1^z, -S_2^z, \, ..., \, -S_L^z \rangle,
\end{equation}

\noindent
where $|S_1^z, S_2^z, \, ..., \, S_L^z \rangle$ labels the spin configurations of the chain; and (ii) space-reflection or parity $\hat P$, with eigenvalues $p=\pm1$, defined as 

\begin{equation}
\hat P |S_1^z, S_2^z, \, ..., \, S_L^z \rangle = | S_L^z, S_{L-1}^z, \, ..., \, S_1^z \rangle.
\end{equation}

Given that $\hat m_z$ commutes with $\hat T$, these two symmetry operations can be used to block-diagonalize Eq.~\eqref{eq:Hamiltonian}. The resulting eigenstates can then be labeled by both the total 
magnetization $m_z$ and the total crystal momentum $K$. 
On the other hand, the discrete symmetries can only be used for block diagonalization in specific subsectors \cite{Sandvik}: 
(i) $\hat Z$ commutes with both $\hat P$ and $\hat T$ but only does so with $\hat m_z$ in the sector with zero total magnetization; 
(ii) $\hat P$ commutes with $\hat m_z$ but does so with $\hat T$ only in the subsectors with zero and $\pi$ total crystal momentum.

Besides this set of symmetry operations, the periodic XXZ chain has additional symmetries at the dense set of commensurate ``roots of unity" anisotropies $\tilde \Delta_{MN} = \cos(\pi M /N)$, 
with $N, M \in \mathbb{Z}^+$ coprimes and $N > M$,  for which there exists a large class of zero-energy $N$-particle ``excitations''  ~\citep{DeguchiFabriciusMcCoy, Braak2001}, in the language of Ref.~\citen{Braak2001}. 
These excitations give rise to degeneracies between a parent state with $m_z= m_z^{\mbox{\scriptsize max}}$ and states with $m_z=m_z^{\mbox{\scriptsize max}}-l \, N$, where $ 0 \le l \le 2 m_z^{\mbox{\scriptsize max}}/N$. 
Deguchi et al.~\citep{DeguchiFabriciusMcCoy} numerically studied these degeneracies and found the corresponding multiplets have multiplicity 

\begin{equation}
\left( \begin{array}{c}
			2 \, m_z^{\mbox{\scriptsize max}}/N \\
			l
\end{array} \right),
\label{eq: degeneracy_commensurate}
\end{equation}

\noindent 
in the commensurable case $m_z = 0 \, \, (\mbox{mod} \, N)$, and 

\begin{equation}
\left( \begin{array}{c}
			2 \, [m_z^{\mbox{\scriptsize max}}/N] + \alpha  \\
			l
\end{array} \right),
\label{eq: degeneracy_uncommensurate}
\end{equation}

\noindent 
in the incommensurable case $m_z \neq 0 \, \, (\mbox{mod} \, N)$, where $[x]$ is the greatest integer contained in $x$, and $\alpha = 0, 1$ or 2. 
In particular, the degeneracies in the commensurable case were related to the $sl_2$-loop algebra~\citep{DeguchiFabriciusMcCoy}, whose generators commute with the XXZ Hamiltonian at roots of unity when $m_z = 0 \, \, (\mbox{mod} \, N)$. 
However, the incommensurable sectors do not have this symmetry.  We shall refer to the quantum symmetry in the full Hilbert space by $\Lambda$. 

\begin{figure}[ttp!]
\centering 
\includegraphics[width=8.0cm]{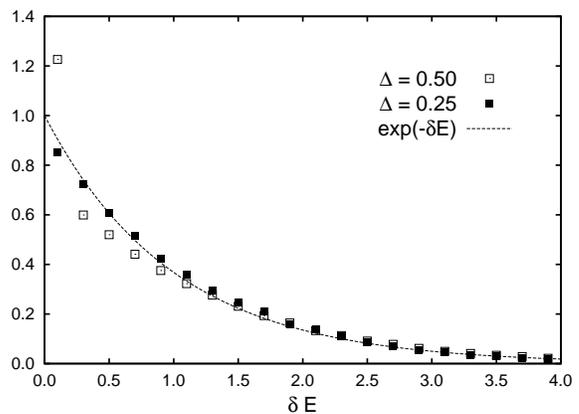}
\caption{XXZ 20-site chain level-spacing distribution for states within the sector of zero total magnetization for two different anisotropies. The distribution is averaged over each spin-inversion sector $z$, for every crystal momentum, as well as over parities for $K=\{0, \pi\}$. Note the peak at $\delta E = 0$ for $\Delta = 0.5$ signaling the presence of additional degeneracies.}
\label{fig: Level-spacing-distr}
\end{figure}

The additional degeneracies in the system at these special $\tilde \Delta_{MN}$ points can be readily verified by inspecting the level spacing distribution~\cite{PhysRevB.68.052510}. 
Indeed, being an integrable model (i.e.~exactly solvable by Bethe Ansatz) the XXZ chain has a macroscopic number of conserved quantities which allow to fully diagonalize the Hamiltonian. 
The spacing distribution of the resulting eigenvalues follows a Poisson distribution~\citep{PoissonStat}. In contrast, as shown in Fig.~\ref{fig: Level-spacing-distr}, at the commensurate anisotropy $\Delta = \tilde \Delta_{1,3} = 1/2$ the level spacing distribution deviates from the Poisson distribution and displays a small peak at $\delta E=0$, signaling the presence of additional degeneracies. 

We compute the spin stiffness (or Drude weight in the equivalent fermionic picture) of the XXZ model through Eq.~\eqref{eq:Drude2}. The ``$z$" component of the current operator entering this equation is given by

\begin{equation}
\hat j^z = \sum\limits_{j =1}^L i \frac{J}{2} \left( \hat S_j^{+} \hat S_{j+1}^{-} - \hat S_j^{-} \hat S_{j+1}^{+} \right),
\label{eq: current}
\end{equation}

\noindent
and it is odd under parity and spin-inversion, and even under translations and spin-rotations, i.e.~$\hat O \hat j^z \hat O = \eta \, j^z$, where $\eta = -1$ for $\hat O = \hat P, \, \hat Z$ and $\eta = 1$ for $\hat O = \hat T, \, \hat m_z$. Hence $\hat j^z$ connects states with the same magnetization $m_z$ and total crystal momentum $K$ but with opposite parities $z$ and $p$.
\begin{table}[ttp!]
\centering
\begin{tabular}{| c | m{4.8cm} |}
\hline
{\bf Symmetry} & {\bf Current carrying states} \\
\hline
$\hat Z$, $\hat P$ and $\Lambda$ & \multirow{2}{4cm}{Degenerate multiplets from $\Lambda$ symmetry only.} \\
$\hat Z$ and $\Lambda$ & \\
\hline
$\hat P$ and $\Lambda$ & All pairwise degenerate states of different parity, including $\Lambda$ multiplets. \\
\hline 
$\hat Z$ and $\hat P$ & \multirow{2}{4cm}{None.} \\
$\hat Z$ & \\
\hline
$\hat P$ & All pairwise degenerate states of different parity. \\
\hline
$\Lambda$ & Every state, including the additional degenerate $\Lambda$ multiplets. \\
\hline
None & All states. \\
\hline
\end{tabular}
\caption{Current carrying states for blocks with fixed $K$ and $m_z$, depending on the presence of spin-inversion $\hat Z$, parity $\hat P$, and quantum $\Lambda$ symmetries. }
\label{table: current-carrying-states}
\end{table}

Finally, to study the role of the discrete symmetries in the evaluation of Eq.~\eqref{eq:Drude2}, we break them using an irrational magnetic flux. The effect of threading the periodic spin chain with such a flux $\phi$ (we chose $\phi = \sqrt{2}$) is incorporated by gauging it into the spin-ladder operators via the usual Peierls substitution $\hat S_j^{\pm}\rightarrow e^{\pm \textrm{i} \phi j/L} \hat S_j^{\pm}$. Thus the flux parameter $\phi$ enters both Eq.~\eqref{eq:Hamiltonian} and \eqref{eq: current}. 

We shall find the presence of the flux also lifts every $\Lambda$-related degeneracy, as we will expatiate more fully in the next section.

\section{Results}

The numerical evaluation of Eq.~\eqref{eq:Drude2} may be performed either in the canonical or grand canonical ensemble. In the former case we sum over eigenstates with a fixed total magnetization $m_z$, 
whereas in the latter the summation includes eigenstates from all the different magnetization sectors. 
In what follows we consider only spin chains with {\it even} $L$.

Let us start by summarizing a couple of empirical observations regarding the many-body states connected by the current operator~\eqref{eq: current} when the Drude weight is evaluated within the canonical ensemble. 
We shall see that when parity and spin-inversion symmetries are broken, as e.g.~in symmetry blocks with $m_z \neq 0$ and $K \neq \{0, \, \pi\}$, or in the presence of the flux $\phi$, {\it all} nondegenerate eigenstates of Eq.~\eqref{eq:Hamiltonian} carry finite currents. %This observation can be understood in terms of Mazur's inequality\citep{Mazur_ineq}: in the absence of spin-inversion the spin current overlaps with the {\it local} even-$\hat Z$ conserved quantities of the model\citep{PhysRevB.55.11029}. The set of nondegenerate energy eigenstates thus spans a subspace in which components of $\hat j^z$ do not decay. 
In particular, and for arbitrary anisotropy $\Delta$, the main contribution to the spin Drude weight comes precisely from these nondegenerate eigenstates. 
In contrast, at roots of unity anisotropies $\tilde \Delta_{MN}$, and in the absence of a flux, we shall find evidence indicating that the set of degenerate eigenstates 
associated with the $\Lambda$ quantum symmetry give the dominant contribution to $D$ in the thermodynamic limit, and these are the only current carrying states if spin-inversion symmetry is present.
Table~\ref{table: current-carrying-states} recaps the main points we discuss below.  

\subsection{\normalsize Current carrying states}
Let us now describe the contribution from various current carrying states in the different magnetization sectors to $D$, and the influence of the flux on these contributions.\\

{\bf (i) $m_z = 0$ sector}: due to the discrete symmetries present in this subsector, and to spin current being odd under their action, only matrix elements from degenerate states of the form 

\vspace{-0.1cm}
$$\langle m, K, z| \hat j^z |-z, K, n \rangle, \quad \mbox{with} \hspace{0.5em}\, K \neq \{0, \pi\}$$ 
\vspace{-0.1cm}
or 
\vspace{-0.1cm}
$$\langle m, K, z, p| \hat j^z |-p, -z, K, n \rangle, \quad \mbox{with} \hspace{0.5em}\, K = \{0, \pi\},$$
\vspace{-0.1cm}
\noindent
and $E_n=E_m$ contribute to the Drude weight, Eq.~\eqref{eq:Drude2}. 
\vspace{0.1cm}

For anisotropy values different from $\tilde \Delta_{MN}$ we find degenerate states only within the sectors of $K=\{0, \pi\}$. 
These states belong to different parity sectors but have the same spin-inversion signature and hence are not connected by the current operator. 
Therefore, there is no net current for any $L$ and the Drude weight vanishes exactly in the sector of zero total magnetization.
Note that at the isotropic point ($\Delta = 1$) the $SU(2)$ symmetry was exploited to show explicitly that $D=0$ for any $\beta$ \cite{CampbellProsen}.

For $\Delta=\tilde \Delta_{MN}$, on the other hand, we find degenerate states due to the additional quantum symmetry only for chains of size 

\begin{equation}
L \ge  L_{\mbox{\scriptsize min}} \equiv 2N,
\end{equation}

\noindent 
the number of which rapidly increases with system size.  
Indeed, the spectral degeneracies related to the $\Lambda$ symmetry are split between sectors of $m_z=0$ (mod $N$)\citep{DeguchiFabriciusMcCoy}. These have multiplicities given by Eq.~\eqref{eq: degeneracy_commensurate}, which implies the existence of $\Lambda$-degenerate states within sectors of fixed magnetization. 
For instance, one can choose $m_z^{\mbox{\scriptsize max}} = N$ and thus find the symmetry sectors with magnetization $m_z=N-lN$ with $0 \le l \le 2$ to be degenerated. 
The smallest possible $\Lambda$-degenerate subspace within a magnetization sector $m_z=0$ (mod $N$) has dimension $2$.
It immediately follows that for system sizes $L = L_{\mbox{\scriptsize min}}=2N$ one should find the first pair of $\Lambda$-degenerate states {\it within the} $m_z=0$ {\it sector}. 
Such a pair is degenerated with the parent state, which has all spins up and magnetization $m_z=N$.

 We have checked numerically this is indeed the case: consider the number of degenerate states within the zero magnetization sector for the two anisotropies $\Delta=0.25$ and $\Delta = \tilde \Delta_{1,6}$, as shown in table~\ref{table: degeneracies}. For $L=12$ and $\Delta = \tilde \Delta_{1,6}$ we find as expected the first two $\Lambda$-degenerate states. Their common eigenvalue equals the eigenvalue of the $m_z=6$ ($=N$) state, i.e.~their parent state. 
 For $L=14$ the $m_z=6$ sector has 14 nondegenerate states. At $\Delta = \tilde \Delta_{1,6}$ each of these states  becomes degenerated with two zero-magnetization states --- hence the 28 additional degenerate states within the $m_z=0$ sector we observed when diagonalising the model (table~\ref{table: degeneracies}). 
Likewise for $L=16$ the $m_z=6$ sector has 113 nondegenerate and 7 degenerate states. Each of these 120 states is degenerated with two zero-magnetization states at $\Delta = \tilde \Delta_{1,6}$ resulting in 240 $\Lambda$-degenerate states. Yet we only see 228 additional degenerate states in table~\ref{table: degeneracies};
disagreements like this are only apparent, and are found every time the parent sector shows some degeneracy~\citep{OnDegeneraciesL16Delta1_6}.

Most importantly, these $\Lambda$-degenerate states are arranged in pairs with opposite spin-inversion signature (as well as opposite parity within the $K=0, \pi$ sectors), and all of them are found to carry finite currents. Hence $\Lambda$-degenerate states alone, occurring only at the dense set of anisotropies $\tilde \Delta_{MN}$,  are responsible for the finite Drude weight at zero-magnetization densities in a canonical ensemble calculation. 

\begin{table}[ttp!]
\centering
\begin{tabular}{| c | c | c |}
\hline
$L$ & deg($\Delta = 0.25$) & deg($\Delta = \tilde \Delta_{1,6}$) \\
\hline
8 & 4 & 4 \\
10 & 20 & 20 \\
12 & 96 & 98 \\
14 & 364 & 392 \\
16 & 1364 & 1592 \\
\hline
\end{tabular}
\caption{Number of degenerate states in the sector of zero total magnetization, for different system sizes and two anisotropy values. Note how the first two degenerate states due to the $\Lambda$ symmetry 
appear at $L=2N=12$.}
\label{table: degeneracies}
\end{table}
\vspace{0.1cm}

{\bf (ii) $m_z \neq 0$ sector}: in the sectors of finite total magnetization there is no spin-inversion symmetry and hence every state potentially carries a current. The relevant matrix elements now are

\vspace{-0.1cm}
$$\langle m, K| \hat j^z |K, n \rangle, \quad \mbox{with} \hspace{0.5em}\, K \neq \{0, \pi\}$$ 
\vspace{-0.1cm}
or 
\vspace{-0.1cm}
$$\langle m, K, p| \hat j^z |-p, K, n \rangle, \quad \mbox{with} \hspace{0.5em}\, K = \{0, \pi\},$$
\vspace{-0.1cm}

\noindent
with $E_n=E_m$. For $\Delta \neq \tilde \Delta_{MN}$ and $K=\{0, \pi\}$ we find pairwise degenerate states of opposite parity, each of which are connected through the spin current operator. 
The rest of the $K$-subsectors have no degeneracies, and all of their (nondegenerate) states are found to carry finite currents.
Since all $K$-sectors are about the same dimension, it follows that nondegenerate current-carrying states yield the dominant contribution to the Drude weight --- an observation which has been made before~\citep{PhysRevB.77.245131}. 
For $\Delta=\tilde \Delta_{MN}$ one finds additional degeneracies in all $K$-subsectors, the number of which rapidly increases with system size. 
Furthermore, all of these additional degenerate states are found to be connected by the spin current operator. 

\begin{figure}[ttp!]
\centering 
\includegraphics[width=8.0cm]{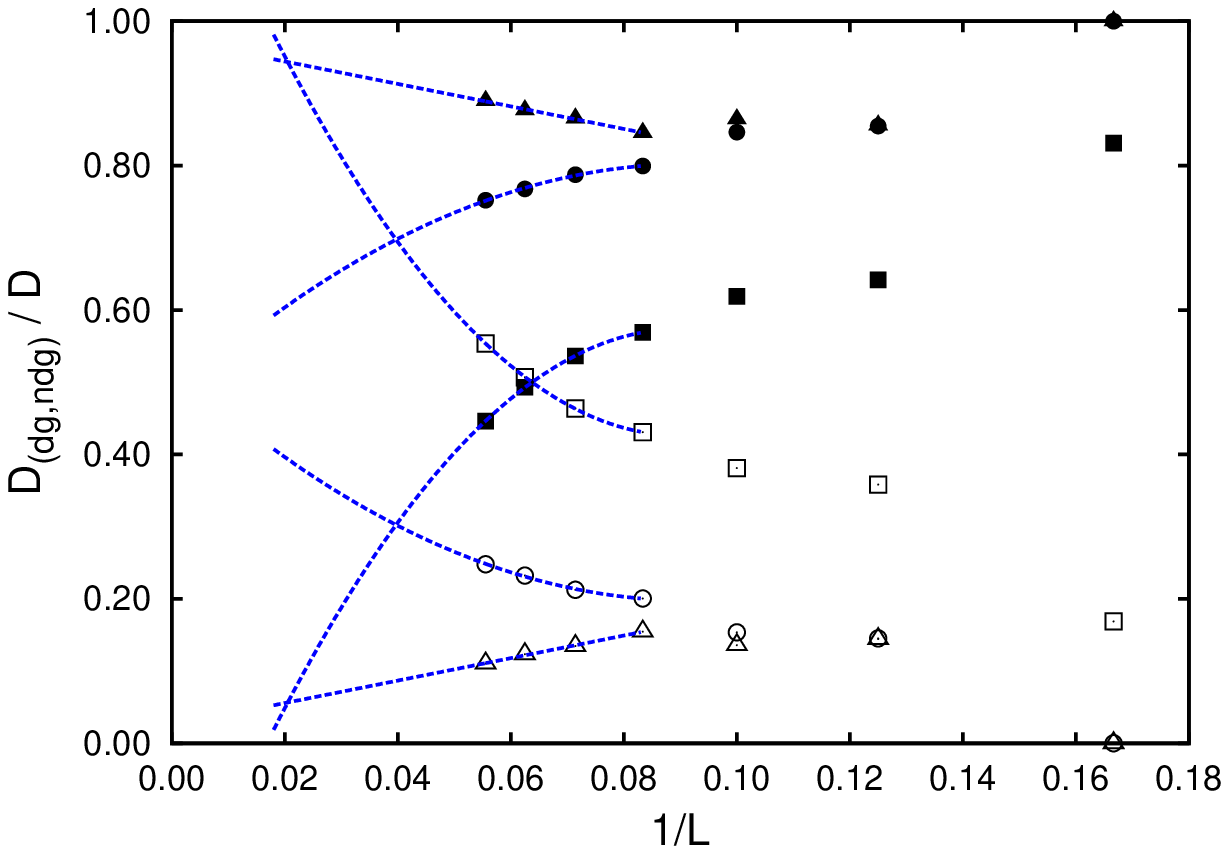}
\caption {Nonfluxed GCE ratios of the contribution to the Drude weight coming from degenerate/nondegenerate states only $D_{\mbox{\scriptsize (dg,ndg)}}$ to the total Drude weight $D$ 
(i.e.~including both degenerate and nondegenerate states). Filled (empty) symbols correspond to nondegenerate (degenerate) contributions. 
Squares indicate data for $\Delta = \tilde \Delta_{1,3}$, circles for $\Delta = \tilde \Delta_{2,5}$ and triangles for $\Delta = 0.25$. 
Triangles also correspond to \textit{any other} incommensurate value or commensurate value with $N > L/2$, as e.g. $\tilde{\Delta}_{13,31} \approx 0.25$.
The dashed lines are second order polynomial fits included to guide the eye. }
\label{fig: ratios}
\end{figure}

Remarkably, from the system sizes and anisotropies $\tilde \Delta_{MN}$ (with $N<L/2$) available to us, $\Lambda$-degenerate states turn out to give the dominant contribution to the Drude weight in the thermodynamic limit when summing over all magnetization sectors --- see the discussion in subsection~\ref{subsec: symmetry_and_ensemble} and Fig.~\ref{fig: ratios}. These observations, together with the fact that only $\Lambda$-degenerate states contribute to $D$ in the sector of zero total magnetization, hint at the fundamental role of these degenerate states in determining the transport properties of the infinite system.

\begin{figure*}[ttp!]
\centering 
\includegraphics[width=15.0cm]{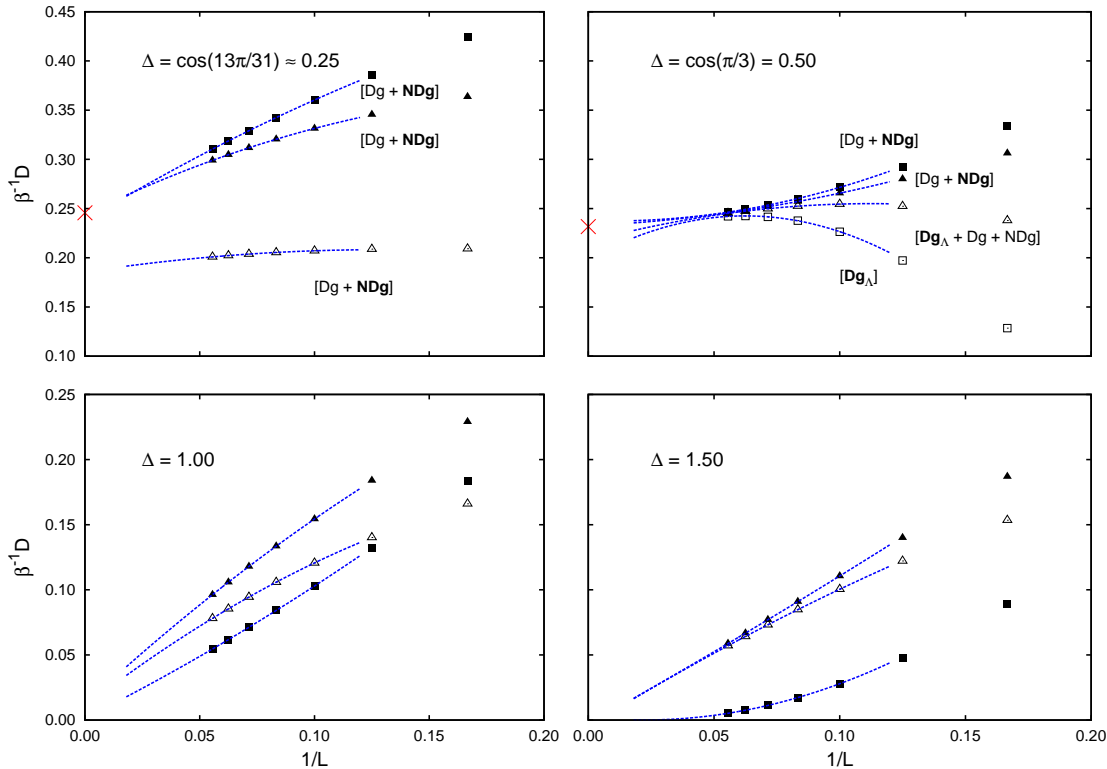}
\caption{Finite size scaling of the Drude weight for different values of the anisotropy $\Delta$, calculated within both canonical (CE) and grand canonical ensemble (GCE), and in the presence and absence of a magnetic flux. The CE computations are carried our for states with zero total magnetization.
Symbol coding: $\blacktriangle$ GCE with flux, $\vartriangle$ GCE without flux, $\blacksquare$ CE with flux and $\square$ CE without flux. The dashed lines correspond to second order polynomial fits and are to 
be taken as guide to eye. 
Top panels show results for the gapless phase, with the crosses representing exact results in the thermodynamic limit \cite{ProsenI, Ilievski}, and the type of current carrying states that contribute to the 
finite-$L$ $D$ values labeled as nondegenerate ($\textrm{NDg}$), degenerate ($\textrm{Dg}$), and $\Lambda$-degenerate states ($\textrm{Dg}_{\Lambda}$); the dominant contributor is highlighted in bold. 
Bottom panels show results for $\Delta \ge 1$, consistent with vanishing Drude weight for $L \rightarrow \infty$, independent of ensemble or symmetry.}
\label{fig:Results}
\end{figure*}

\vspace{0.1cm}
{\bf (iii) $m_z = 0$ sector and finite magnetic flux}: if $\Delta \neq \tilde \Delta_{MN}$ and $K=\{0, \pi\}$ we find the same number of degenerate states as for the nonfluxed model. 
In this case however, since parity and spin inversion are broken, degenerate states are only connected to themselves (i.e.~$m=n$) by the current operator, and give, together with the nondegenerate states,  finite contributions to the summation in Eq.~\eqref{eq:Drude2}.
For all the other momenta one finds that all states are nondegenerate and carry finite currents. Summarizing, the relevant matrix elements in this case are
$$\langle m, K| \hat j^z |K, n \rangle \, \delta_{mn}.$$
Now if $\Delta = \tilde \Delta_{MN}$  we find the presence of the flux lifts all the extra degeneracies coming from the additional quantum symmetry. Hence, in sharp contrast to the nonfluxed cases, the spin Drude weight does not have singular contributions at any special set of anisotropies, and thus is found to be finite within the entire gapless phase. 
This suggests $D$ might be a continuous function of $\Delta$ when the model is pierced by a magnetic flux on a finite ring. 

\vspace{0.1cm}

{\bf (iv) $m_z \neq 0$ sector and finite magnetic flux}: in this case we find no degenerate state in any of the symmetry subsectors, and that all the (nondegenerate) states therein carry finite currents regardless of $\Delta$. All contributing matrix elements, within any $K$-sector,  are therefore of the form
$$\langle m, K| \hat j^z |K, n \rangle \, \delta_{mn}.$$
\vspace{-1.0cm}

\subsection{\normalsize Drude weights anomalies: symmetries and ensemble}
\label{subsec: symmetry_and_ensemble}

We present infinite temperature Drude weight results in Fig.~\ref{fig:Results}, evaluated within the grand canonical (GCE) and canonical ensemble (CE) in the $m_z = 0$ sector, both with and without a flux, 
for even-length chains of size $L=6-18$. We also show a second order $(1/L)$-polynomial fit to the data points of $L=10-18$ to guide the eye.

Per usual statistical mechanics GCE and CE should be equivalent in the thermodynamic limit, especially if the quantity measured is a meaningful one in this limit~\cite{PhysRevB.77.161101}. 
We will find that although in the gapped phase $D$ quickly becomes independent of both ensemble and symmetries as $L$ increases, in the gapless phase the presence of spin-inversion symmetry makes convergence 
towards the thermodynamic limit remarkably slow, depending on how fast the $\Lambda$ degeneracies start showing up upon increasing system size. 
\vspace{0.1cm}

{\bf Gapless phase. ---} For $\Delta < 1$ it is rigorously known that the high temperature transport is ballistic~\cite{ProsenGapless} i.e.~$D > 0$ in \textit{any} magnetization sector, 
especially the zero magnetization sector where usual local conserved quantities alone (together with Mazur's inequalities) do not settle the issue. 
However computing the actual value of $D$ is a different matter, with $D$ lower-bounded more strictly recently~\cite{ProsenI}, and even claimed to be exactly computable~\cite{Ilievski}. 
The explicit expression for the bound at the $\tilde{\Delta}_{MN}$ points reads

\begin{equation}
\label{eq: lbound}
\beta^{-1} D \ge \frac{\pi}{8} \frac{\sin^2(\pi M/N)}{\sin^2(\pi/N)} \left(1-\frac{N}{2\pi} \sin(2\pi/N)\right),
\end{equation}

\noindent 
which agrees with the Thermodynamic Bethe Ansatz (TBA) result~\cite{Zotos, Benz, HerbrychZotos} 

\begin{equation}
\beta^{-1}D = \pi\frac{\gamma - \sin{(2\gamma)/2}}{8\gamma},
\label{eq: ThermodynamicBetheAnsatz}
\end{equation}

\noindent
with $\gamma = \cos^{-1}\Delta$, at the anisotropies $\tilde \Delta_{1N}$.

In the top panels of Fig.~\ref{fig:Results} we show the finite size data for $\beta^{-1} D$ as a function of system size in the gapless phase, for the two anisotropies $\Delta = \tilde \Delta_{13,31}$ and $\tilde \Delta_{1,3}$. 
In both cases we note (i) $D$ always extrapolates in the thermodynamic limit to a finite value $\mathcal{O}(1)$ and (ii) the fluxed cases, for finite $L$, always gives larger $D$ values. 
The second point is analogous to the observation that in the GCE odd-length chains have larger $D$ values than even-length chains\cite{CabraHonecker}. 
The reason behind such a difference lies mainly in the presence/absence of spin-inversion symmetry (for odd-length chains spin inversion is always broken). 
Indeed, as pointed out above, in the absence of spin inversion all eigenstates, whether degenerate or not, of the XXZ Hamiltonian are connected by the current and give a finite contribution to $D$. 
If the states have a definite $z$-signature, in contrast, many of the eigenstates of $H$, as e.g. all states with $m_z=0$ for $\Delta \neq \tilde \Delta_{MN}$, carry no current. 
Hence, for finite systems $D$ assumes greater values in the GCE for either fluxed or odd-length periodic chains. 

Let us focus on $\Delta =\tilde  \Delta_{1,3} = 0.5$ first. We see that independent of keeping or breaking the discrete symmetries and the type of ensemble chosen, the Drude weight values always agree with one another, 
even for moderately small $L \approx 18$.  
Pivotally, they tend to extrapolate to the same value in the thermodynamic limit which is in good agreement with the lower bound~\eqref{eq: lbound} --- marked with a cross in the figure. 

For the second case $\Delta = \tilde \Delta_{13,31} \approx 0.25$ the results seem generally the same 
but distinctly different in one aspect: the finite-size data for the GCE without flux seems to extrapolate to a different thermodynamic limit from the fluxed GCE or CE cases. 
The latter two here tend to extrapolate to the same thermodynamic limit in good agreement with the lower bound, Eq.~\eqref{eq: lbound}. 
Note also the CE nonfluxed data is absent, for it is exactly zero at these system sizes. 
We understand the source of the discrepancy between nonfluxed GCE/CE and fluxed GCE/CE $-$ as well as to theoretical predictions $-$ as follows: 
we know from the previous section that for $\Delta = \tilde \Delta_{13,31}$ the degeneracies due to the $\Lambda$ symmetry start showing up for spin chains of length $L \ge 2 \times 31=62$. 
We also know that these degenerate states, the number of which rapidly increases with system size, carry finite currents and thus should play some role in determining the thermodynamic limit of $D$.
 In fact only these states contribute to $D$ in the zero magnetization sector, whereas they give the largest contribution to the GCE calculation of $D$ in the thermodynamic limit. 
To illustrate this last point we computed, within the nonfluxed GCE, ratios between the contributions to Eq.~\eqref{eq:Drude2} from nondegenerate (degenerate) states and the total Drude weight, labeled with filled (empty) symbols in Fig.~\ref{fig: ratios}, as a function of system size. 
The squares corresponds to data for $\Delta = \tilde \Delta_{1,3}$ and clearly show that the contribution from degenerate states quickly starts dominating the Drude weight upon increasing system size. In fact $\Lambda$ degeneracies are already present for chains of length $L \ge 6$, and these increase from 2 for $L=6$ to more than 36794 current-carrying states for $L=18$ in the zero magnetization sector alone, which amounts to more than 75\% of the total number of states within that sector. 
The subsequent dominance of degenerate $\Lambda$ states over nondegenerate states thus explains why, in the top panels of Fig.~\ref{fig:Results}, both nonfluxed GCE and CE data approach the lower bound from below upon increasing system size.
In contrast the triangles in Fig.~\ref{fig: ratios}, which correspond to the incommensurate anisotropy $\Delta = 0.25$, display a completely different behavior, namely nondegenerate states always dominate the GCE calculation of the Drude weight.

Remarkably, data points for other $\Delta \neq \tilde \Delta_{MN}$ or $\tilde \Delta_{MN}$ with $L < 2N$ (as e.g.~$\tilde \Delta_{13,31}$) --- that is, data points with no $\Lambda$ symmetry --- are found to overlap with the triangles in Fig.~\ref{fig: ratios}. 
Indeed the circles in this same figure, which label data for $\tilde \Delta_{2,5}$, overlap with the triangles for $L<2N = 10$ whereas for $L \ge 2N=10$ the contribution from the $\Lambda$ degeneracies becomes finite and increases its relative weight in $D$ with increasing system size --- 
clearly, the actual contribution of the $\Lambda$-degenerate states is simply the difference between circles (or squares for $\tilde \Delta_{1,3}$) and triangles. 
Therefore, finite spin chains with no $\Lambda$ symmetry display the same ratio of the Drude weight computed from nondegenerate states to the total Drude weight, regardless of the anisotropy.

One can thus firmly speculate that (i) in order to see the GCE/CE nonfluxed data for $\tilde \Delta_{13,31}$ extrapolate to the exact result in Fig.~\ref{fig:Results}, the current-carrying $\Lambda$-degenerate states need to be taken 
into account, which amounts to considering chains of length $L > 64$, and that (ii) these states are precisely the ones contributing to the Drude weight in the thermodynamic limit. 
Unfortunately, system sizes as $L > 64$ are out of the question when all exact eigenstates are needed. 
Nevertheless these observation do suggest the fundamental role played by $\Lambda$-degenerate states in contributing to the thermodynamic spin Drude weight. 
This is particularly clear from the nonfluxed CE results for $\tilde \Delta_{1,3}$ which agree in the thermodynamic limit with Eq.~\eqref{eq: lbound} and for which we know only $\Lambda$-degenerate states contribute. The reader will find additional data for different $\tilde \Delta_{MN}$ points 
supporting this claim in the Appendix. 

\begin{figure}[ttp!]
\centering 
\includegraphics[width=8.0cm]{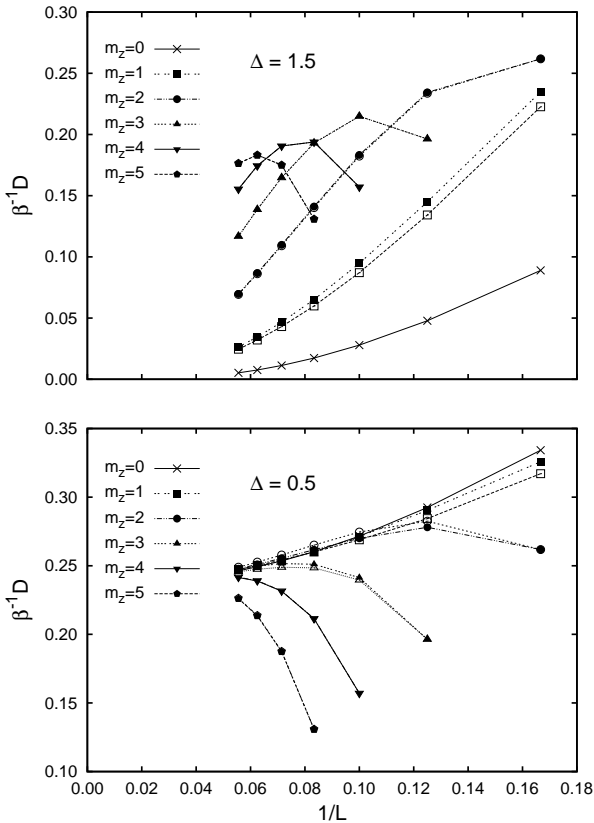}
\caption{Finite size scaling of the Drude weight in the canonical ensemble for two $\Delta$ values, 
calculated in the presence (full symbols) and absence (empty symbols) of a magnetic flux, for different magnetization sectors.  Crosses indicate $m_z=0$ values in the presence of a flux. As $m_z$ and $L$ increase the effect of flux disappears for both anisotropies; note however that for larger $\Delta$ the $m_z = 0$ states contribute the least, whereas it is the opposite for smaller $\Delta$. This  explains why, when the weighted average is taken, $m_z = 0$ CE data is above (below) the GCE data as seen in the upper (lower) right panel of Fig.~\ref{fig:Results} for these two anisotropies. Note that our $D$ computation is for fixed $m_z$ (rather than $m_z/L$) and hence the tendency to extrapolate to zero, as well as the nonmonotonicity at large $L$ seen in the top figure.}
\label{fig:Example}
\end{figure}

We close this subsection with the following open questions and remarks: first, consider $\Delta = 0.25$ for which the $\Lambda$ symmetry is not present. For the system sizes considered here the absolute value of the difference $D_L[\Delta=0.25]-D_L[\tilde \Delta_{13,31}]$ is always of the order of $10^{-4}$ or less, and so the data for $\Delta = 0.25$ is indistinguishable (not shown) from that plotted in the top-left panel of Fig.~\ref{fig:Results}. Due to the absence of the $\Lambda$ symmetry, however, the Drude weight computed within the nonfluxed CE vanishes for all $L$. We may then ask whether $D$ computed within the nonfluxed GCE will actually extrapolate to zero in the thermodynamic limit at {\it exactly} $\Delta=0.25$, as well as at any other $\Delta \neq \tilde \Delta_{MN}$, as to agree with the canonical ensemble result.  
Second, we noted above that in the presence of a flux all $\Lambda$-related degeneracies are lifted. In such cases no slow convergence is found for {\it any anisotropy} and both CE and GCE fluxed data always extrapolate to the same finite value in the thermodynamic limit. This seems to imply $D$ is a continuous function of $\Delta$ in the presence of a flux.
Third, we have noticed that upon increasing the anisotropy, the contribution to the Drude weight from the symmetry sectors of larger magnetization densities slowly start dominating the summation in Eq.~\eqref{eq:Drude2}, as exemplified by Fig.~\ref{fig:Example} for $\Delta = 1.5$ (top panel), whereas the trend is exactly the opposite for the smaller anisotropy $\Delta = 0.5$ (bottom panel). 
This makes the nonfluxed GCE computation for large $\Delta$ less susceptible to the physics of the zero- and low magnetization sectors, and hence to the symmetry-related anomalies.
In fact, we have carried out computations for the large anisotropies $\tilde \Delta_{1,6} \approx 0.87$ and $\tilde \Delta_{1,8} \approx 0.92$ and confirmed that, even when the current-carrying $\Lambda$-degenerated states have yet to appear, the nonfluxed GCE data extrapolates to the same thermodynamic limit as the fluxed GCE and CE data 
(see e.g.~the lower left panel in Fig.~\ref{fig:Appendix} in the Appendix).

\vspace{0.2cm}

{\bf Isotropic point and gapped phase. ---} Away from the gapless phase the high-temperature Drude weight  is known to vanish, as follows from the spin-reversal invariance of the thermodynamic macrostates sustained by the system at these anisotropies~\cite{Ilievski}, and as predicted from numeric simulations~\cite{CabraHonecker, MarkoXXZ, MarkoGapped, ZnidaricVarma}.
We confirm this picture holds for GCE and CE, again with and without a flux, in the bottom panels of Fig.~\ref{fig:Results}. 

For $\Delta \ge 1$ the $\Lambda$ symmetry is absent~\cite{DeguchiFabriciusMcCoy, Braak2001} and the nonfluxed CE data with $m_z=0$ case gives once more identically zero $D$ due to spin-inversion 
(and hence it is not shown). 
Accordingly, the fluxed GCE results, which include finite contributions from the zero magnetization sector, are above the non-fluxed ones. 

\begin{figure*}[ttp!]
\centering 
\includegraphics[width=18cm]{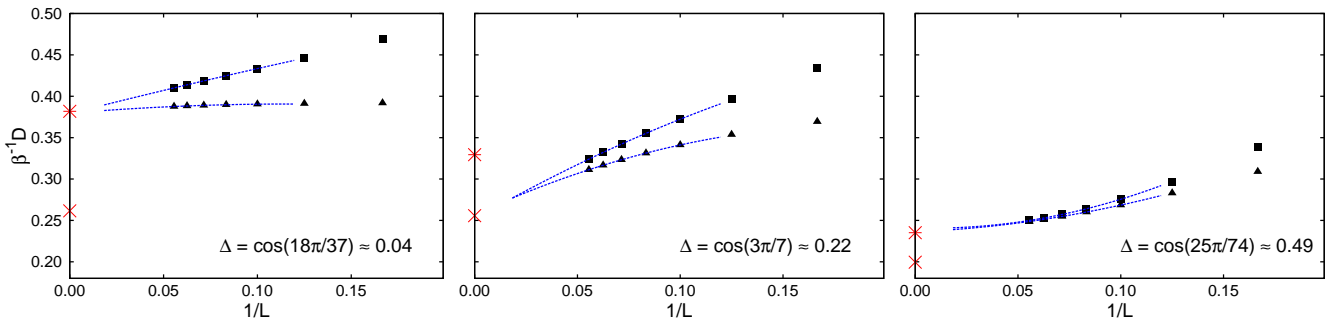}
\caption{Finite size scaling of the Drude weight computed in both fluxed CE ($\blacksquare$) and GCE ($\blacktriangle$) within the gapless phase. The crosses represent the exact lower bound results, Eq.~\eqref{eq: lbound}, and the stars the Thermodynamic Bethe Ansatz results, Eq.~\eqref{eq: ThermodynamicBetheAnsatz}.}
\label{fig: Extrapol_data}
\end{figure*}

The main difference one notes here, in contrast to the two gapless points from above, is that the fluxed CE $D$ values are lower than those for the GCE (both with and without a flux). 
This is so because, as already mentioned, for these anisotropies and available system sizes, the contributions from the symmetry sectors of finite magnetization densities dominate the summation in Eq.~\eqref{eq:Drude2} (see Fig.~\ref{fig:Example}).
The GCE results thus show less dependence on flux, and hence on $\hat Z$ and $\hat P$ upon increasing $\Delta$, as was already seen in Fig.~\ref{fig:Results}.

\subsection{\normalsize Extrapolation anomalies within the gapless phase}
%\subsection*{EXTRAPOLATION ANOMALIES WITHIN THE GAPLESS PHASE}
\vspace{0.25cm}

Despite seemingly good convergence of the numerical results to the lower bound of Ref.~\citen{ProsenI} when all symmetries are broken --- that is, in the presence of the flux ---  we will now see that there are caveats. 

Herbrych et al.~\cite{HerbrychZotos} found that for anisotropies $\Delta > 0.5$ the Drude weight generically scales as $1/L$ and extrapolates to the TBA result, Eq.~\eqref{eq: ThermodynamicBetheAnsatz}. 
This result is intriguing for we know Eq.~\eqref{eq: ThermodynamicBetheAnsatz} is only valid at the set of anisotropies $\tilde \Delta_{1, N}$, and so one expects, from our previous analysis as well as the results in Ref.~\citen{ProsenI}, 
$D$ at $\tilde \Delta_{MN}$ might generically extrapolate to Eq.~\eqref{eq: lbound} instead. 
They also pointed out that for some anisotropies $\Delta < 0.5$, the Drude weight does not scale as $1/L$ due to finite-size low-frequency contributions in the regular part of the Kubo conductivity, which lead to a finite correction $\delta D$ in the thermodynamic limit. 
It was found, in particular, that when such a correction was {\it not} taken into account, the Drude weight data did not extrapolate to Eq.~\eqref{eq: ThermodynamicBetheAnsatz}. 

We complement these observations by reporting that, for a set of anisotropies in the region ($\Delta < 0.5$) where the low-frequency anomalies were found, and if no correction $\delta D$ is considered, $D$ {\it for fluxed rings} extrapolates to the lower bound, Eq.~\eqref{eq: lbound}, as was already seen in the top left panel of Fig.~\ref{fig:Results} for $\tilde \Delta_{13,31} \approx 0.25$. The center panel of Fig.~\ref{fig: Extrapol_data} showing results for $\tilde \Delta_{3,7} \approx 0.22$ together with the additional data set $\tilde \Delta_{4,9} \approx 0.17$, $\tilde \Delta_{2,5} \approx 0.31$ and $\tilde \Delta_{3,8} \approx 0.38$ in the Appendix further support this observation.

Interestingly enough, as shown in the left and right panels of Fig.~\ref{fig: Extrapol_data}, when approaching either the point $\tilde \Delta_{1,2} = 0$ or $\tilde \Delta_{1,3}=0.5$ the finite-size data extrapolates instead to the TBA solution, Eq.~\eqref{eq: ThermodynamicBetheAnsatz}. 
Additional calculations at commensurate anisotropies $\tilde \Delta \approx 0.1$ and $\tilde \Delta \approx 0.4$ seem to extrapolate to $D$ values lying {\it in between} Eq.~\eqref{eq: ThermodynamicBetheAnsatz} and the lower bound. 
Such a crossover between these two analytic results must be a finite-size effect. 

At the moment we are unable to theoretically account for the convergence to these two distinct results, apart from highlighting 
that the low-frequency anomalies of Ref.~\citen{HerbrychZotos} seem to vanish when close to the points $\tilde \Delta_{1,2}$ and $\tilde \Delta_{1,3}$, i.e.~precisely when the data extrapolates to Eq.~\eqref{eq: ThermodynamicBetheAnsatz}. 
We have also run simulations for different $\tilde \Delta_{MN} > 0.5$ in order to further compare our results with those of  Ref.~\citen{HerbrychZotos}. However our naive extrapolation does not allow us to clearly discern whether the data extrapolates to Eq.~\eqref{eq: ThermodynamicBetheAnsatz} or to the lower bound, 
because the numerical values of these two results start getting closer upon increasing $\Delta$.

\section{Summary and discussion}
\begin{figure*}[ttp!]
\centering 
\includegraphics[width=18.0cm]{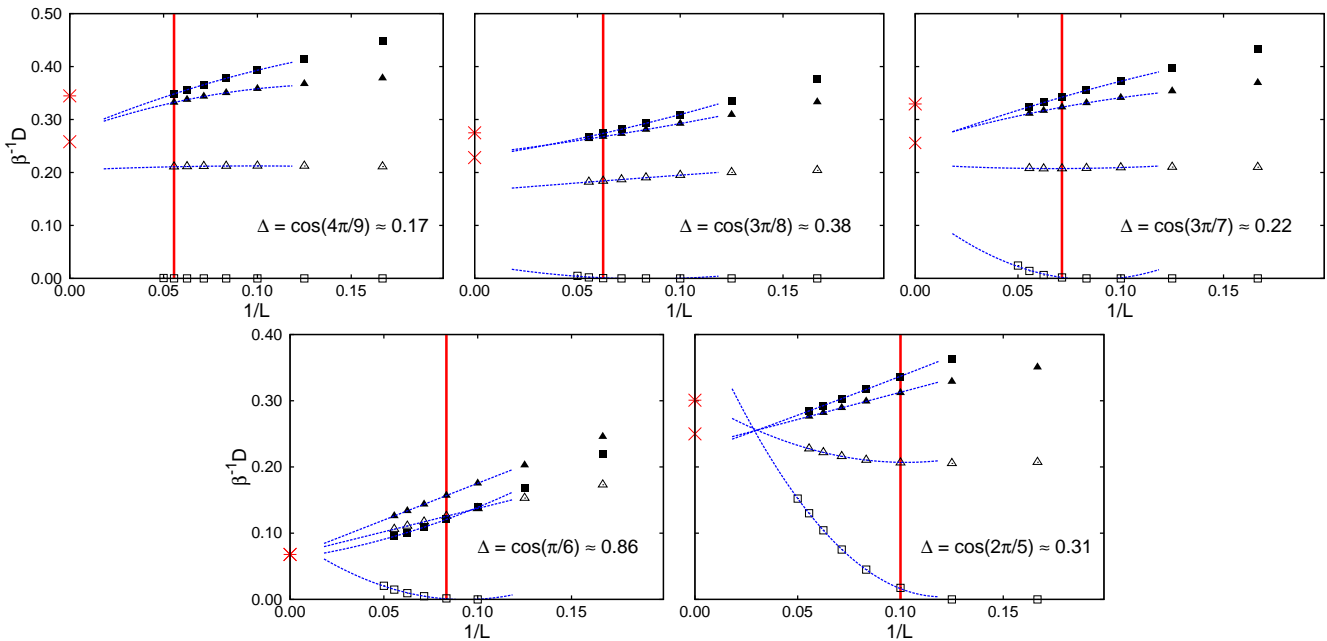}
\caption{Finite size scaling of the Drude weight for additional commensurate anisotropies $\tilde \Delta_{MN}$, ordered by decreasing $N$ and computed within both canonical (CE) and grand canonical ensemble (GCE), and in the presence and absence of a magnetic flux. 
Symbol coding: $\blacktriangle$ GCE with flux, $\vartriangle$ GCE without flux, $\blacksquare$ CE with flux and $\square$ CE without flux. 
The dashed lines correspond to second order polynomial fits and are to be taken as guide to eye. 
The red vertical lines indicate the system size at which the $\Lambda$ degeneracies start taking place.
The crosses represent the exact lower bound results, Eq.~\eqref{eq: lbound}, and the stars the Thermodynamic Bethe Ansatz results, Eq.~\eqref{eq: ThermodynamicBetheAnsatz}.
}
\label{fig:Appendix}
\end{figure*}
We studied the influence of symmetries --- namely parity, spin-inversion and $\Lambda$ symmetries --- and ensemble (canonical or grand canonical) on the high temperature spin Drude weight in the anisotropic XXZ chain. 
Introducing a flux provides a convenient way to break these symmetries and to study their effect upon the numeric evaluation of the Drude weight. 

At arbitrary (incommensurate) anisotropies, for which the $\Lambda$ symmetry is absent, we found that the dominant contribution to the (finite-size) Drude weight comes from nondegenerate states. 
In particular, we found that degenerate states do not carry any current in the presence of spin inversion symmetry. 
The latter observation allowed us to identify the finite-size discrepancies between GCE computations for even- and odd-length chains reported in the literature as consequence of the absence/presence of 
spin-inversion symmetry in the model.

In contrast, at commensurate ``roots of unity" anisotropies $\Delta = \cos(\pi M/N)$ with $N$, $M \in \mathbb{Z}^+$ coprimes and $N>M$, additional degeneracies associated to the underlying $\Lambda$ symmetry 
start showing up for system sizes $L \ge 2N$, where $N$ can be large depending on the anisotropy. 
Remarkably, from the system sizes and commensurate anisotropies (with $N<L/2$) available to us,  we find these degenerate states --- the number of which rapidly increases with system size --- 
give the dominant contribution to $D$ in the thermodynamic limit, and these are the only current-carrying states if spin-inversion is present. 
Thus, in the canonical ensemble and for states with zero total magnetization, $D$ is only finite at the dense set of commensurate anisotropies $\tilde \Delta_{MN}$ for $L \ge 2N$.
These findings seem to support the recently found (fractal) structure of $D$ within the gapless phase of the model. 
It remains an interesting open question how to reconcile the observed relevance of $\Lambda$-degenerate eigenstates of the model in contributing to the thermodynamic Drude weight and the Thermodynamic Bethe Ansatz calculation, i.e.~the classification of the model's eigenstates into high-weight and descendant states of e.g.~the $sl_2$-loop symmetry (regular Bethe eigenvectors of the XXZ model at roots of unity have been shown to be highest weight vectors of the $sl_2$-loop algebra in some restricted $m_z$ sectors~\citep{BetheStatesHighestWeight}).  We leave this classification for future work. 

In the presence of a flux spin-inversion is broken and the $\Lambda$ degeneracies were found to be lifted. 
In this case the spin Drude weight does not have singular contributions at any special set of anisotropies, which suggests $D$ might be a continuous function of $\Delta$ when the model is pierced by a magnetic flux on a finite ring. 
We found that within the gapless phase ($\Delta < 1$) the finite-size Drude weight extrapolates to either the recently computed lower-bounds or to the TBA solution, Eq.~\eqref{eq: ThermodynamicBetheAnsatz}, independent of the ensemble used. In particular, convergence towards Eq.~\eqref{eq: ThermodynamicBetheAnsatz} is found whenever the previously reported low-frequency anomalies\citep{HerbrychZotos} are absent.

We confirm absence of ballistic transport for $\Delta \geq 1$, both in the grand canonical ensemble and the zero magnetization sector, with or without the discrete symmetries preserved. 

Finally, we remark that with increasing anisotropy the contribution to the Drude peak from the zero magnetization sector goes from being the most dominant to the least dominant. Given that only this magnetization sector is affected by spin-inversion symmetry, the role of the latter decreases with increasing the anisotropy in a grand canonical ensemble calculation. 

These zero-frequency anomalies show up, in addition to new ones, in the corresponding finite-frequency and finite-momentum response functions too, and will therefore be directly relevant to experimental probes. 
This is reported in a follow-up work \cite{to.be.published}. \\

We thank V. Oganesyan for related collaboration, and B. Doyon, F. Heidrich-Meisner, T. Prosen, V. Rittenberg, and M. \v Znidari\v c for discussions, and the latter for helpful comments on an earlier version of the manuscript. 
One of us (RJS) acknowledges financial support from the H2 branch of the Bonn-Cologne Graduate School of Physics and Astronomy.

\appendix

\section{Gapless phase: additional data points}

Figure~\ref{fig:Appendix} shows the finite size scaling of the spin Drude weight for additional commensurate anisotropies $\tilde \Delta_{MN}$, computed within both canonical (CE) and grand canonical ensemble (GCE), and in the presence and absence of a magnetic flux. 
The panels are ordered by decreasing $N$, or equivalently by the increasing number of system sizes with the $\Lambda$ symmetry (i.e. the $L > 2N$ limit). 
The vertical red lines mark the system size $L=2N$ for which the first pair of $\Lambda$ degeneracies shows up. 

Two points are noteworthy: first, note how when decreasing $N$ convergence of nonfluxed CE and GCE data improves. This follows because the number of current-carrying $\Lambda$ states increases with our 
available system sizes, i.e. as the number of data points on the left of the vertical red line increases, it becomes conspicuous that the effect of the ensemble under consideration becomes less vital.
Second, note how all the fluxed data sets extrapolate well to the lower bound, Eq.~\eqref{eq: lbound}, marked with a cross in the figure. These additional results support our claim in the main text 
that for $\tilde \Delta_{MN} < 0.5$ the finite-size data for fluxed chains generically extrapolates to the lower bound.

\bibliographystyle{unsrt}
\bibliography{Ref6}

\begin{thebibliography}{10}

\bibitem{Kohn64}
W.~Kohn.
\newblock {\em Phys. Rev.}, 133:A171--A181, 1964.

\bibitem{CastellaZotos}
H.~Castella, X.~Zotos, and P.~Prelovsek.
\newblock {\em Phys. Rev. Lett.}, 74:972, 1995.

\bibitem{ZotosPrelovsek}
X.~Zotos and P.~Prelovsek.
\newblock {\em Phys. Rev. B}, 53:983, 1996.

\bibitem{PhysRevB.55.11029}
X.~Zotos, F.~Naef, and P.~Prelovsek.
\newblock {\em Phys. Rev. B}, 55:11029--11032, 1997.

\bibitem{Mazur_ineq}
P~Mazur.
\newblock {\em Physica}, 43:533, 1969.

\bibitem{MillisAndrei}
B.~N. Narozhny, A.~J. Millis, and N.~Andrei.
\newblock {\em Phys. Rev. B}, 58:R2921, 1998.

\bibitem{CabraHonecker}
F.~Heidrich-Meisner, A.~Honecker, D.~C. Cabra, and W.~Brenig.
\newblock {\em Phys. Rev. B}, 68:134436, 2003.

\bibitem{PhysRevB.77.245131}
S.~Mukerjee and B.~S. Shastry.
\newblock {\em Phys. Rev. B}, 77:245131, 2008.

\bibitem{HerbrychZotos}
J.~Herbrych, P.~Prelovsek, and X.~Zotos.
\newblock {\em Phys. Rev. B}, 84:155125, 2011.

\bibitem{PhysRevLett.108.227206}
C.~Karrasch, J.~H. Bardarson, and J.~E. Moore.
\newblock {\em Phys. Rev. Lett.}, 108:227206, 2012.

\bibitem{PhysRevB.87.245128}
C.~Karrasch, J.~Hauschild, S.~Langer, and F.~Heidrich-Meisner.
\newblock {\em Phys. Rev. B}, 87:245128, 2013.

\bibitem{PhysRevLett.112.120601}
R.~Steinigeweg, J.~Gemmer, and W.~Brenig.
\newblock {\em Phys. Rev. Lett.}, 112:120601, 2014.

\bibitem{VarmaSanchez}
V.~K. Varma and R.~J. S\'anchez.
\newblock {\em Phys. Rev. A}, 92:013618, 2015.

\bibitem{Karrasch2017}
C.~Karrasch.
\newblock {\em New J. Phys.}, 19:033027, 2017.

\bibitem{PhysRevB.94.201112}
M.~Filippone, P.~W. Brouwer, J.~Eisert, and F.~von Oppen.
\newblock {\em Phys. Rev. B}, 94:201112, Nov 2016.

\bibitem{ProsenI}
T.~Prosen and E.~Ilievski.
\newblock {\em Phys. Rev. Lett.}, 111:057203, 2013.

\bibitem{Ilievski}
Enej Ilievski and Jacopo De~Nardis.
\newblock {\em Phys. Rev. Lett.}, 119:020602, 2017.

\bibitem{to.be.published}
R.~J. S\'anchez, V.~K. Varma, and V.~Oganesyan.
\newblock To be published.

\bibitem{Sandvik}
A.~W. Sandvik.
\newblock {\em AIP Conf. Proc}, 1297:135, 2010.

\bibitem{DeguchiFabriciusMcCoy}
T.~Deguchi, K.~Fabricius, and McCoy~B. M.
\newblock {\em J. Stat. Phys.}, 102:701, 2001.

\bibitem{Braak2001}
D.~Braak and N.~Andrei.
\newblock {\em Journal of Statistical Physics}, 105(3):677--709, 2001.

\bibitem{PhysRevB.68.052510}
K.~Kudo and T.~Deguchi.
\newblock {\em Phys. Rev. B}, 68:052510, 2003.

\bibitem{PoissonStat}
D.~Poilblanc, T.~Ziman, J.~Bellisard, F.~Mila, and G.~Montambaux.
\newblock {\em EPL}, 22:537, 1993.

\bibitem{CampbellProsen}
J.~M.~P. Carmelo, T.~Prosen, and D.~K. Campbell.
\newblock {\em Phys. Rev. B}, 92:165133, 2015.

\bibitem{OnDegeneraciesL16Delta1_6}
For $L=16$ the $m_z=6$ sector has 113 nondegenerate and 7 degenerate states,
  the latter with the same eigenvalue $\varepsilon$ and labeled by the same
  total momentum $K$. On the one hand every nondegenerate $m_z=6$ state is
  degenerated with two $m_z=0$ states, which yields 226 $\Lambda$-degenerate
  states. On the other hand we do find 14 $\Lambda$-degenerate states with
  eigenvalue $\varepsilon$ in the zero magnetization sector. We noted
  nevertheless that at $\Delta = \tilde \Delta_{1,6}$ six different eigenvalues
  from 12 pairwise degenerate $m_z=0$ states (of different parity and same
  spin-inversion signature within the $K=0$ sector, thereby not contributing to
  $D$) become equal to $\varepsilon$. Hence the difference of 228 degenerate
  states in table~\ref{table: degeneracies}.

\bibitem{PhysRevB.77.161101}
M.~Rigol and B.~S. Shastry.
\newblock {\em Phys. Rev. B}, 77:161101, 2008.

\bibitem{ProsenGapless}
T.~Prosen.
\newblock {\em Phys. Rev. Lett.}, 106:217206, 2011.

\bibitem{Zotos}
X.~Zotos.
\newblock {\em Phys. Rev. Lett.}, 82:1764, 1999.

\bibitem{Benz}
J.~Benz, T.~Fukui, A.~Kluemper, and C.~Scheeren.
\newblock {\em J. Phys. Soc. Jpn. Supp.}, 74:181, 2005.

\bibitem{MarkoXXZ}
M.~Znidaric.
\newblock {\em Phys. Rev. Lett.}, 106:220601, 2011.

\bibitem{MarkoGapped}
M.~Znidaric.
\newblock {\em Phys. Rev. B}, 90:115156, 2014.

\bibitem{ZnidaricVarma}
M.~Znidaric, A.~Scardicchio, and V.~K. Varma.
\newblock {\em Phys. Rev. Lett.}, 117:040601, 2016.

\bibitem{BetheStatesHighestWeight}
T.~Deguchi.
\newblock {\em J. Phys A: Math. Theor.}, 40:7473, 2007.

\end{thebibliography}

\end{document}